# Calibration of clinical trial sample size based on design utility

Charu Gupta[1], Gabriel Innocenzi[2], Christina Yap[3], Daniel Jackson[4], Fabio Rigat[1*]

## Abstract

Clinical trial design relies on both statistical and clinical considerations for pre-specification of potentially practice-changing target treatment effects. As larger trials tend to be associated with high power and modest minimal detectable benefit, trial sample size is typically calibrated with reference to relevant precedents to prevent overpowering. Albeit trial sponsors and regulators are accustomed to this practice, there is scope for simplification to enhance the robustness, transparency and cross-trial consistency of this aspect of the design process. To this end, a design utility index is proposed here as a formal basis for sample size calibration, balancing the increase in power at higher sample sizes against the corresponding reduction in the magnitude of minimum detectable treatment effects, without requiring additional statistical assumptions or bespoke software. Application of utility calibration to a broad range of designs demonstrates consistency with regulatory expectations for minimum power, particularly in confirmatory settings, and effective protection against overpowering and against overly aggressive interim analyses.

**Author affiliations**:

1. AstraZeneca Oncology Biometrics, Gaithersburg, USA and Cambridge, UK.
2. AstraZeneca External Expert Engagement Services, Lisbon, Portugal.
3. Group Leader, ICR-CTSU Early Phase and Adaptive Trials, UK
4. AstraZeneca Oncology Biometrics Statistical Innovation, Cambridge, UK.

* Corresponding author: fabio.rigat@astrazeneca.com

## Introduction

The harmonised principles guiding clinical trial design [1-4] require trial protocols and statistical analysis plans to define the key design components including target population, variables to be measured, primary measure of interest and handling of intercurrent events. These components motivate and justify the quantitative decision rules determining the primary analysis outcome, including for designs using multiple endpoints and sequential data analyses [5-8]. In this context, trial sample size is typically set to achieve 80%-90% power, subject to the false positive error probability not exceeding 5% for registrational trials. These statistical properties of the primary analysis outcome determine the smallest statistically significant treatment effect, commonly referred to as the critical value (CV). As the trial sample size increases, statistical power increases while CVs become more modest. Consequently, when a trial yields a statistically positive outcome under high-power and modest CV conditions, it is well known that a misalignment may arise between statistical significance and its clinical relevance [9-12].

The outcomes of two recent oncology clinical trials in gastric or gastroesophageal junction (gastric/GEJ) adenocarcinoma, KEYNOTE-585 [13] and REGARD [14], illustrate circumstances where differences in sample size, statistical significance and clinical benefit ultimately resulted in different regulatory outcomes within the same indication. KEYNOTE-585 failed to show statistical significance for the primary event free survival (EFS) endpoint (HR = 0.81, 804 participants) at the pre-specified p-value boundary, although reporting clinically relevant EFS and overall survival (OS) median improvements of respectively 10.5- and 2.7-months OS, and no regulatory approvals ensued. Conversely, REGARD (355 participants) secured regulatory approval after reporting a statistically significant overall survival HR of 0.776 with modest 1.4-months improvement in median OS. These examples illustrate a context where trials may fail despite reporting outcomes within clinically meaningful ranges when statistical significance is diluted by design complexity, and they may instead succeed despite reporting modest benefit when powered to detecting relatively small treatment effects. Additional oncology case studies are reported in the Supplementary Materials.

Recognising in full the challenges of this contextual complexity, the design utility proposed here provides a reference framework enabling a formal consideration of current mainstream statistical design criteria together with their associated minimal statistically significant treatment benefit. Beyond promoting consistency between statistical and clinical trial design criteria, this approach also addresses a broader call to accelerating pragmatic innovation of clinical trial design in the pursuit of more efficient options, as crisply expressed in [15]: "*Statistical methods in clinical research tend to become entrenched. Innovations threaten the status quo. The "right way" becomes frozen in lore. This is so even when the "right way" is not best. "Statistical significance" and the associated requirement of "high power" is an example. This attitude is an impediment to efficient design*". The design utility illustrated here relates to these concerns by providing a calibration framework balancing the increase in power at higher sample sizes against the corresponding reduction in the magnitude of minimum detectable benefit.

Utility-based experimental design is not a new concept [16-27], but no utility measures are routinely considered yet as a basis for optimisation of pivotal clinical trial design. To this end, Section 1 builds the definition of design utility on more general utility formulations developed from the 1920s onwards [28-30]. Section 2 shows that application of design utility to sizing common clinical trial designs does not require additional assumptions nor implementation of non-standard statistical methods, demonstrating its practical potential on a broad basis. Section 3 focuses in on the application of design utility to calibrating the sample size of group-sequential designs with time to event primary endpoints, which are commonly used in late phase clinical development in oncology. First, it is shown that utility-based sample size calibration based provides an effective mechanism to discourage overly aggressive interim analyses. Second, selecting the trial sample size maximising the final analysis utility is shown to prevent overpowering, providing CVs within a clinically meaningful range and broadly consistent with current best practice. These results are summarised in the discussion, which also highlights potential avenues to further extending the application of utility theory to clinical trial design.

## 1. What is a well powered study? Insufficiency of statistical operational characteristics

Let the motivating trial hypothesis be a statement $H_1$ and let $H_0$ be its null hypothesis, with $H_0$ being rejected by an appropriate statistically positive primary trial outcome "$+_N$" for a trial with total sample size $N$. The dichotomous trial outcome $+_N$ is derived from a pre-specified test statistic and statistical significance threshold [31]. The power associated with $+_N$ is the probability of observing a positive trial outcome under $H_1$, represented here as $p(+_N|H_1)$. The false positive error rate is the probability of observing a statistically positive trial outcome under $H_0$ and is represented as $p(+_N|H_0)$. For randomised trials, sample size is typically calculated to achieve a target power while controlling the false positive error rate at a pre-specified level $0 < \alpha < 1$. These operational characteristics determine the positive likelihood ratio (PLR) of the primary analysis outcome,

$$PLR_N = p(+_N|H_1)/\alpha, \qquad (1)$$

which quantifies how much more likely a positive outcome will be observed under $H_1$ than under $H_0$. For instance, a study designed at $p(+_N|H_1) = 90\%$ and $p(+_N|H_0) = 5\%$ provides $PLR_N = 0.9/0.05 = 18$, showing that observing this positive outcome is 18 times more likely if $H_1$ is assumed to be true compared to when $H_0$ is assumed true. When $H_0$ and $H_1$ are a priori taken to be equally likely, observing this positive outcome demonstrates that $H_1$ is 18 times more likely to be true than $H_0$ [32]. As power generally increases with the sample size $N$ for a fixed false positive error $\alpha$, (1) is a non-decreasing function of $N$. Hence, maximising (1) with respect to $N$ is insufficient to determine an optimal sample size short of the operationally feasible maximum $N_{max}$. This limitation of the statistical operational characteristics of trial outcomes is in practice problematic, because it is not conducive to efficient trial designs and it requires additional considerations to determine trial samples sizes smaller than $N_{max}$. To this end, the minimum detectable benefit associated to a positive trial outcome is usually considered alongside to power and false positive error rate [33-35]. For instance, when the measure of interest is the hazard ratio (HR) estimated from a time to event endpoint, the minimum detectable benefit is

$$MB_N = 1 - HR_{CV}(N), \qquad (2)$$

where the critical value $HR_{CV}(N)$ is the largest statistically significant hazard ratio. Similar definitions of (2) apply to any study design, with appropriate modifications in the measure of interest and of its value under $H_0$, as demonstrated below. Quantification of $MB_N$ for non-randomised designs relies on a fixed value of the efficacy endpoint under the relevant standard of care, typically derived from historical data. Examples of (2) for randomised and non-randomised designs are provided in the examples below, including Bayesian designs [36-38].

### 1.1 Definition of design utility

Insufficiency of the positive likelihood ratio alone as a basis for determining a clinically meaningful sample size was noted above. Equivalently, interpreting the positive likelihood ratio as a measure of design utility cannot yield a utility maximising sample size lower than the maximum feasible $N_{max}$, due to the monotonicity of (1) in $N$. To overcome this limitation, the design utility measure proposed here leverages the opposite behaviours of (1) and (2) with respect to their common design attribute $N$ using the multiplicative functional form:

$$U_N := PLR_N{}^{\gamma} \times MB_N{}^{\delta}. \qquad (3)$$

where $\gamma, \delta > 0$ are fixed weights modulating the relative utility contributions of (1) and (2). For any given value of $\gamma$ and $\delta$, maximising (3) over the range of operationally feasible sample sizes provides a reference solution $N^* := argmax_N\{U_N, N_{min} \leq N \leq N_{max}\}$ for sample size calibration. In all applications presented below the weights $\gamma = \delta = 1$ are used, meaning that power (1) and minimum detectable benefit (2) are taken as equally relevant to determining $N^*$.

The design utility (3) belongs to the family of Cobb-Douglas utility functions [39], which first described an empirically validated model of a two-factors production function later generalised for optimisation of consumer choices, and then studied in the multi-attribute decision analysis literature [40]. In this context, utility functions are maximised subject to a budget constraint representing the total cost of goods or services. In the context of clinical trial design, (3) is maximised with a respect to sample size, which is the common attribute determining both power and minimum detectable benefit, with no need for further constraints and without requiring any additional statistical or clinical assumptions.

Of note, when $\gamma = \delta = 1$ equation (3) is also interpretable as expected utility, normalised by the false positive error rate $\alpha$. Expectation here is taken with respect to the value of the study outcome under the design hypothesis $H_1$, with the conservative minimum benefit $MB_N$ being associated to a positive study outcome and with zero benefit associated to a negative outcome. Hence, when $\gamma = \delta = 1$ the reference sample size $N^*$ maximises the expected design utility.

The main assumption underpinning multiplicative utilities is mutual utility independence [41]. For (3) this means that higher power is always preferred at any value of minimum detectable benefit, and conversely greater minimum detectable benefit values are always preferred at any power level. Under mutual utility independence only the additive, multiplicative or multi-linear forms fulfil the axioms of von Neumann and Morgenstern defining cardinal utility functions [42-43]. The multiplicative form (3) is used here for sample size calibration, noting that its logarithm is an additive utility which can be alternatively used, providing equivalent results in all examples illustrated below.

In all examples provided here, the opposite behaviours of power and of minimum detectable benefit in $N$ result in a concave profile of (3) in sample size. Hence, in these examples the utility-maximising $N^*$ is unique. Also, [40] showed that concave utility functions encode risk aversion. For (3), this means that intermediate sample sizes will tend to be preferred compared to the extremes $N_{min}$ and $N_{max}$. A relevant exception is presented below, in relation to aggressive interim analyses of time to event endpoints at low information fraction.

## 2 Application of design utility to single-analysis trial designs

Application of (3) to early and late clinical trials designs are presented next. An interactive web application implementing the design utility calculations for the following examples is freely accessible at https://gs-intersect.vercel.app/ without registration. This application accepts the design parameters described in the examples below, calls a base R application programming interface (API) supplemented by the *gsDesign* [44] and *clinfun* [45] packages, to calculate the design utility across the feasible sample size range, and returns interactive utility curves, highlighting the utility maximizing design. R scripts for all examples in this paper, including the Bayesian simulations not implemented in the web application, together with the full application source code, are publicly available at https://github.com/gainnoce/gs-intersect.

### 2.1 Single arm design, continuous endpoint

The design hypothesis $H_1$ of the single arm proof of concept study considered here is that exposure to an investigational treatment will improve the average disease activity score (DAS28) endpoint used in clinical development of therapeutic agents against rheumatoid arthritis [46-47]. Under $H_1$ the average DAS28 is assumed to be equal to 6 at baseline and 4 at week 12, so that the target effect size is $6 - 4 = 2$. At design stage, the difference in DAS28 between baseline and week 12 for each study participant is modelled as a Gaussian random variable with standard deviation $\sigma = 3$, so that the standardised effect size (Cohen's $d$) equals $d = (6-4)/3 = 0.67$. The paired Z statistic is an appropriate statistical model of $d$ here when the variance of the measure of interest is fixed, and the paired Student-t test statistic is appropriate when the variance of the measure of interest is unknown. Using these common statistical models, the minimum detectable benefit (2) is defined here as the smallest statistically significant average difference in DAS28 calculated among study participants, which is $MB_N = Z_{1-\alpha/2} \times \frac{\sigma}{\sqrt{N}}$ for the Z test and $MB_N = T_{1-\frac{\alpha}{2},(N-1)} \times \frac{\hat{\sigma}}{\sqrt{N}}$ for the Student-t test, respectively when $\sigma$ is fixed or when its estimate $\hat{\sigma}$ is used for data analysis. The values $Z_{1-\frac{\alpha}{2}}$ and $T_{1-\frac{\alpha}{2},(N-1)}$ are respectively the $(1-\alpha/2)$th quantiles of the standard Normal distribution and of the standard Student-t distribution with $(N-1)$ degrees of freedom. When $\gamma = \delta = 1$, the design utility (3) for the paired Z test, $U_{N,\sigma}$, and that for the paired Student-t test $U_{N,\hat{\sigma}}$, are respectively

$$U_{N,\sigma} = \frac{1-\Phi_n\left(Z_{1-\frac{\alpha}{2}}-d\sqrt{N}\right)+\Phi_n\left(-Z_{1-\frac{\alpha}{2}}-d\sqrt{N}\right)}{\alpha} \times Z_{1-\alpha/2} \times \frac{\sigma}{\sqrt{N}}, \quad (4)$$

$$U_{N,\hat{\sigma}} = \frac{1-\Phi_t\left(T_{1-\frac{\alpha}{2},(N-1)}, d\sqrt{N}\right)+\Phi_t\left(-T_{1-\frac{\alpha}{2},(N-1)}, d\sqrt{N}\right)}{\alpha} \times T_{1-\frac{\alpha}{2},(N-1)} \times \frac{\hat{\sigma}}{\sqrt{N}}, \quad (5)$$

where $\Phi_n(x)$ is the standard Normal cumulative distribution function evaluated at the value $x$ and $\Phi_t(x, y)$ is the non-central Student-t cumulative distribution function evaluated at $x$ with non-centrality parameter $y$.

Figure 1 depicts the positive likelihood ratio (1) on the left panel, the critical value (centre) and the design utilities (4) and (5) respectively as dark blue and light blue curves, plotted against sample size $N$ (horizontal axes). The estimated standard deviation $\hat{\sigma}$ is taken to be equal to the true the standard deviation $\sigma$, without loss of generality. The left and central panels in Figure 1 show that the positive likelihood ratio and the minimum detectable benefit respectively increase and decrease with increasing sample size, with the Student-t test having lower power than the $Z$-test at all sample sizes, as expected. The right panel in Figure 1 shows that, although (4) and (5) are not amenable to straightforward analytical maximisation in $N$, the higher power and larger positive likelihood ratio values associated with larger sample sizes are balanced by the associated decrease in minimum detectable benefit, so that their product (3) exhibits a concave profile in $N$, highlighting that the utility is primarily driven by gain in power initially but becomes increasingly driven by a decreasing minimum detectable effect as $N$ increases. Hence, further increases in $N$ contribute relatively little to additional evidential strength while the minimum detectable benefit continues to decrease. The design utilities (4) and (5) identify unique utility-maximising sample sizes of respectively $N$ =17 and $N$ =19 response evaluable study participants under the paired Z and Student-t test statistics, with approximately 80% power and minimal detectable DAS28 average improvement of approximately 1.4 units in both cases. Should clinical investigators find the design characteristics associated with maximum design utility insufficient, additional input on the clinical relevance of the minimum detectable benefits associated with different sample sizes would be needed to further refine the sample size determination.

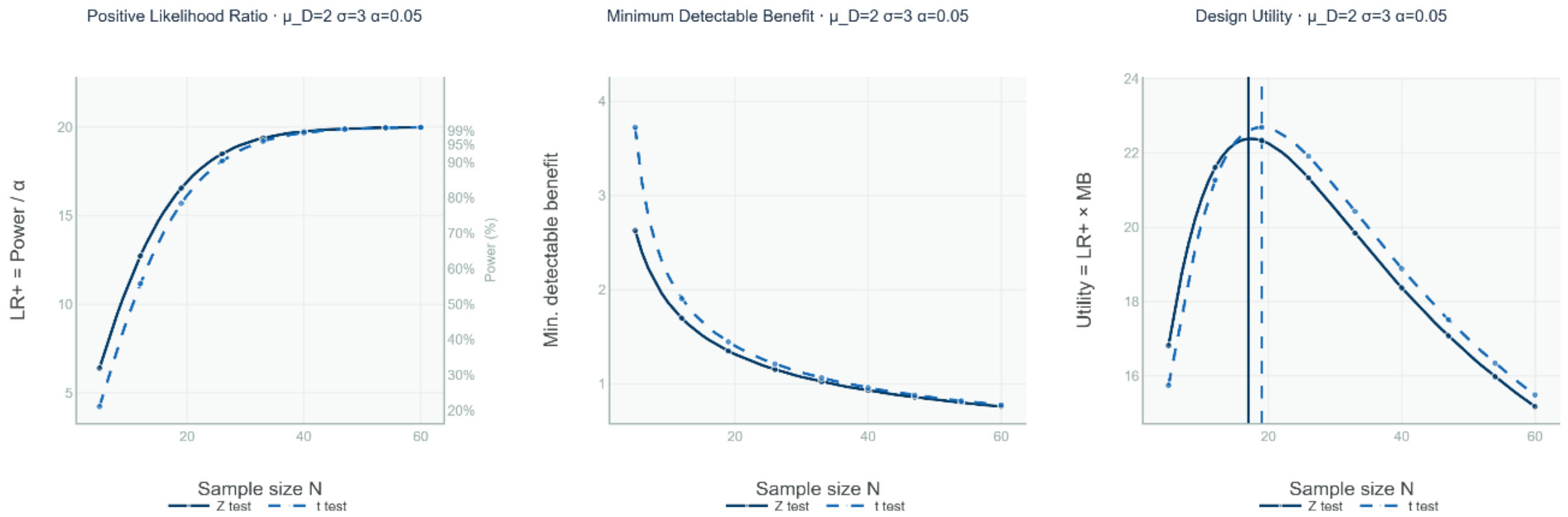

*Figure 1: power and positive likelihood ratio at 5% false positive error rate (left panel), critical value (centre) and design utility (right panel) versus sample size (horizontal axes) for the paired Z test (in dark blue) and the corresponding Student-t -test (in light blue) statistics. The design utility identifies unique utility maximising sample sizes when using either test statistic, with approximately 80% power.*

## 2.2 <u>Simon 2- stage design</u>

Simon 2-stage design supports clinical development decisions pertaining to progression of investigational treatments towards pivotal trials based on single cohort early efficacy readouts, minimising the expected trial sample size at given error probabilities, lower reference and target efficacy levels [48-51]. Decisions under Simon 2-stage are articulated over two pre-planned analyses: an interim futility analysis and a final go-no go analysis. Design utility is applied here to the final analysis of a Simon 2-stage design with binary endpoint and false positive error rate fixed at 5%. The minimum detectable benefit $MB_N$ is defined as the difference between the smallest objective response rate (ORR) associated with a positive final analysis and the lower reference ORR. The null hypothesis here is that the ORR is no greater than the lower reference value 30% and the target ORR response rate is 50%.

The left, central and right panels in Figure 2 depict respectively the positive likelihood ratio (1), the minimum detectable benefit (2) and the design utility (3) of the optimal Simon 2-stage design minimising the expected sample size, plotted against $N$. Figure 2 shows that, although the positive likelihood ratio and the minimum detectable benefit are not smooth functions of sample size due to reliance on the discrete Binomial distribution, the utility maximising sample size $N = 45$ is identified, with associated power of 84% and minimum detectable ORR increase of 10%. Table 1 summarises the sample size maximising the Simon 2-stage design utility (3), the associated power and minimum ORR associated with a positive study outcome over a broader range of alternative hypotheses. Similarly to Figure 2, in all cases a unique global utility maximum was identified, suggesting broad applicability of (3) to this design.

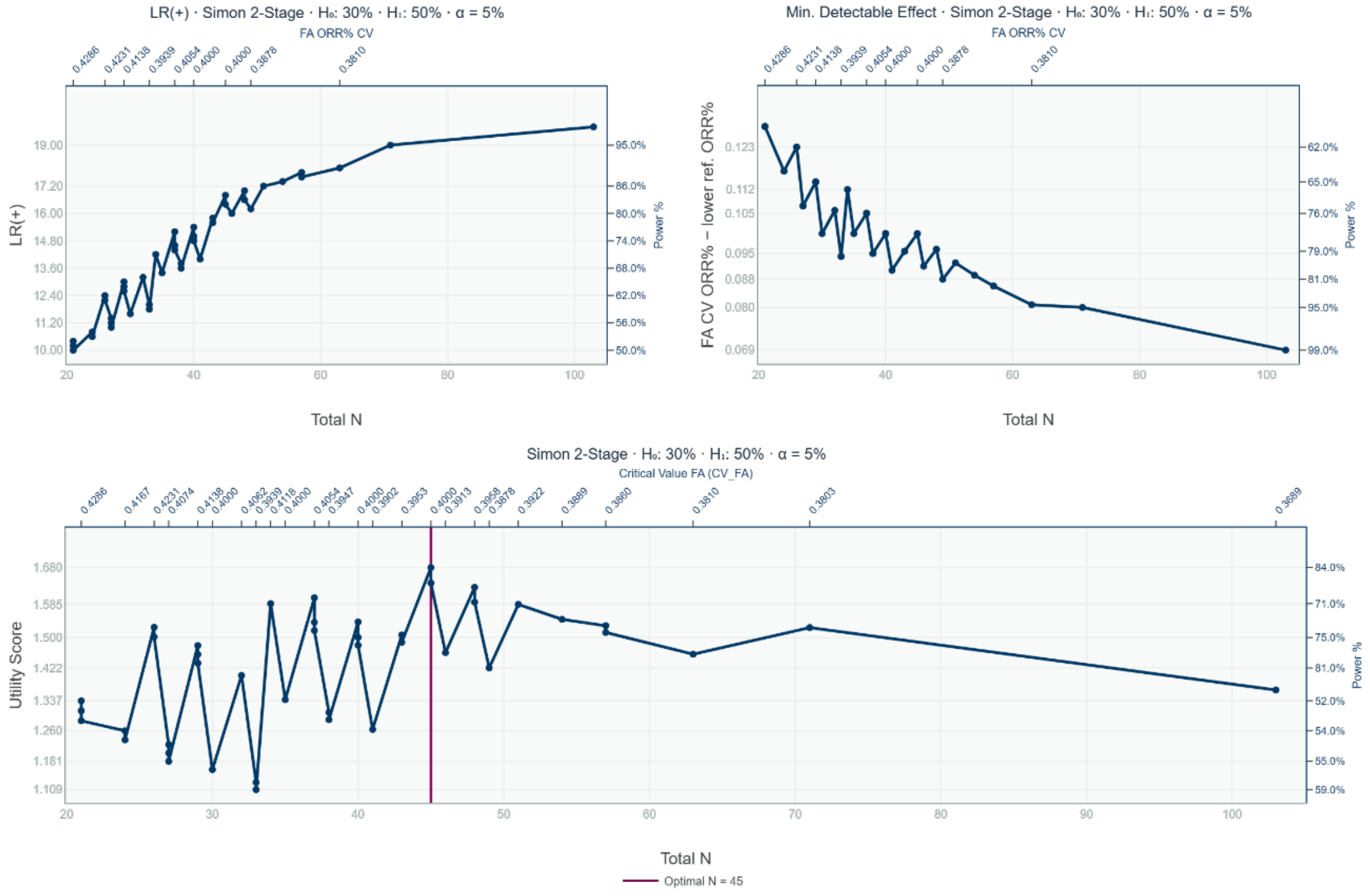

*Figure 2: positive likelihood ratio (top left), minimum detectable benefit (top right) and design utility (bottom) for the Simon 2-stage design assuming 30% lower reference ORR, 50% target ORR and 5% false positive error rate. Design utility (3) has a global maximum at sample size N=45, associated with 84% power. The minimum ORR associated with a positive study outcome at this sample size is 40%.*

| Reference ORR | Target ORR | Max utility $N^*$ | Power | Min positive ORR (Min benefit) |
|---|---|---|---|---|
| 30% | 40% | 248 | 95% | 34.7% (4.7%) |
| | 50% | 45 | 84% | 40% (10%) |
| | 60% | 12 | 63% | 50% (20%) |
| 50% | 60% | 165 | 79% | 55.8% (5.8%) |
| | 70% | 55 | 90% | 60% (10%) |
| | 80% | 23 | 90% | 65.2% (15.2%) |

*Table 1: optimal Simon 2-stage designs maximising design utility at 5% false positive error rate under three alternative hypotheses. A unique design utility maximum is identified in all cases, with associated power ranging between 63% and 90%.*

### 2.3 Difference between proportions - randomised design, single analysis

Figure 3 illustrates the application of design utility to a randomised proof of concept study evaluating the difference in ORR between a standard of care arm and a parallel investigational treatment arm. Analogously to the Simon 2-stage design presented above, the control arm ORR is assumed to be 30% at design stage and the ORR in the investigational arm is 50% under $H_1$. The efficacy measure of interest defining the primary study outcome is the difference in ORR between the investigational arm versus control. False positive error is set at 5% and the same number $N$ of study participants are to be randomised to each study arm.

Under the Gaussian approximation to the probability distribution of the ORR difference, the minimum detectable benefit (2) is quantified by the smallest ORR difference defining statistical superiority $MB_N = Z_{1-\frac{\alpha}{2}} \times \sigma(0.3,0.5,N)$ with standard deviation $\sigma(0.3,0.5,N) = \sqrt{(0.3 \times 0.7 + 0.5^2)/N} = \sqrt{0.46/N}$. The left, central and right panels in Figure 3 depict respectively the positive likelihood ratio (1), the minimum detectable benefit (2) and the design utility (3) for this randomised design, plotted against sample size. Design utility exhibits a unique global maximum at sample size $N$ = 89 response evaluable study participants per study arm, achieving approximately 80% power under $H_1$ and CV of approximately 14% ORR difference in favour of the investigational arm compared with standard of care.

Figure 4 compares power and false positive error calculated under the Gaussian approximation with those obtained using Bayesian inference under the same design hypotheses. The probability of response in each of the study arms is treated as an unknown random parameter with uniform prior probability distribution over the range 0-100%. To estimate Bayesian power and false positive error probabilities, fifty thousand synthetic trial datasets were generated using the Binomial distribution to model the number of responders in each of the study arms under the assumed 30% and 50% ORRs. For each simulated dataset, fifty thousand draws were generated from the Beta posterior distribution of the ORR in each study arm to calculate the Bayesian power. Power here was defined as the proportion of simulated trials where more than 97.5% of the ORR values sampled from the posterior distribution of the investigational arm were greater than the associated ORR values sampled from the posterior distribution of the control arm. Analogous trial simulations were run using identical response rates of 30% in both arms to estimate the Bayesian two-sided false positive error as the proportion of trials simulated under $H_0$ where the posterior probability of the ORR in the investigational arm being greater than that of the control arm exceeded 97.5% or was smaller than 2.5%. Figure 4 shows that Bayesian power closely tracks the Gaussian approximation at all sample sizes, and that the Bayesian false positive probability is accurately and consistently controlled at 5%, confirming that the same value of design utility (3) and the same utility-maximising sample size obtain under this Bayesian model and under the Gaussian approximation for this design. The Bayesian analysis presented in this section is not implemented in the web application to reduce the computational overheads of exact Monte Carlo simulations, which are implemented in the R supplementary script available at https://github.com/gainnoce/gs-intersect.

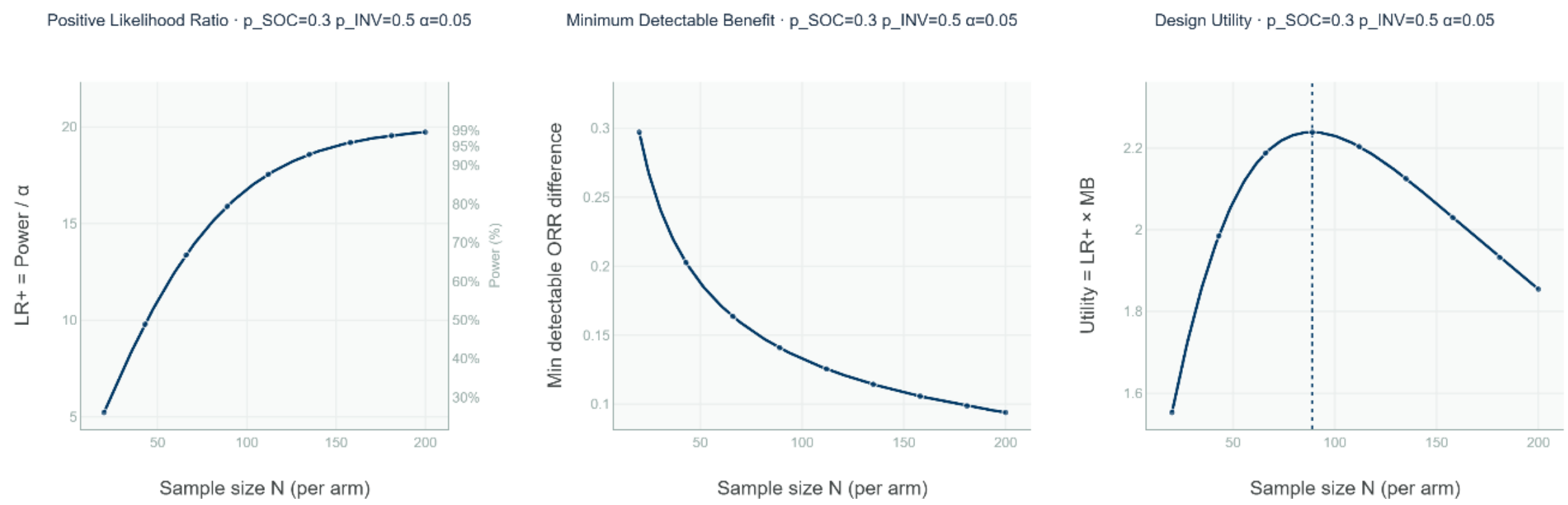

*Figure 3: sample size determination for a parallel 2 arm study with binary efficacy endpoint. The measure of interest is the difference in the proportion of responders between the parallel study arms, assuming the same sample size in both arms. The Gaussian approximation to the probability distribution of the ORR difference is used to calculate the positive likelihood ratio and power (left panel), minimum detectable benefit (central panel) and design utility (right panel) over a range of sample sizes. The design utility identifies a unique global maximum at N =89 participants per arm, with 79.5% power and a critical value of 14.1% difference in response rates in favor of the investigational arm.*

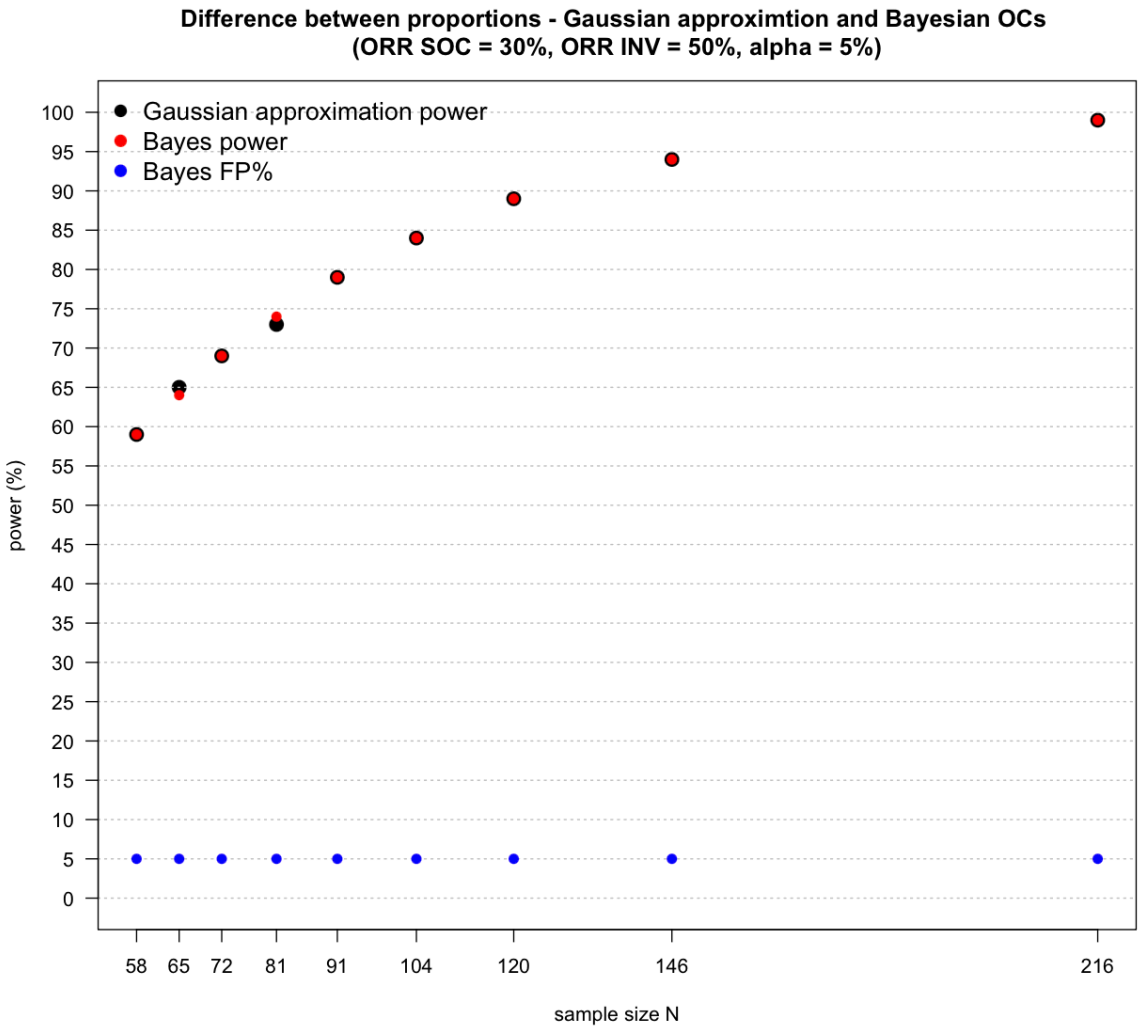

*Figure 4: probability of correctly rejecting the null hypothesis (in black) for the randomised design testing for an ORR difference between two parallel trial arms, plotted against sample size in each of the trial arms, compared with Bayesian power (in red) under ORR uniform priors in both trial arms and with Bayesian false positive probability (in blue). Approximating Bayesian power by Monte Carlo simulation from the exact posterior distributions of the ORR in each trial arm resulted in minor numerical discrepancies from the power values calculated under the Gaussian approximation to the distribution of the ORR difference, based on fifty thousand simulated trial datasets. Further reduction of these numerical errors can be achieved by increasing the number of simulations as needed. As Bayesian power and false positive error probability coincide with those associated to the Gaussian approximation, the design utility and the utility-maximising sample size in both cases takes the same value.*

## 3 Application of design utility to sample size calibration for group-sequential randomised controlled designs with time to event primary endpoint

Group sequential (GS) design [52-54] encompasses a family of statistical methods ensuring control of the family-wise error rate (FWER) over a pre-planned set of analyses. FWER control is typically achieved by pre-specifying how the overall false positive error is partitioned among successive analyses, with early interim analyses being typically allocated a small proportion of the overall false positive probability. As GS designs are typically based on achieving a target power under FWER control, the clinical relevance of hazard ratio critical values associated with a statistically positive outcome is carefully considered for sample size determination. Maximisation of design utility (3) in this context offers a formal framework to support GS design optimisation by balancing the precision of the primary study outcome against the associated minimum treatment benefit.

To demonstrate the practical application of utility-based sample size calibration to GS design, Figure 5 shows the values of (3) plotted against the corresponding number of events expected to be observed at a single interim analysis (IA, on the left) and final (FA) analysis. The median time to event primary endpoint is assumed to be 12 months under the standard of care arm. The target hazard ratio is set at 0.7, which is associated to a median time to event of approximately 17 months under exponential distributions in both trial arms. Enrolment is assumed to complete within 18 months from study start, with expected enrolment rates of 2.5, 5 and 7.5 patients over the first 2,4 and 6 months respectively and 10 patients per month over the next 12 months. The minimum follow-up time for all patients is set at 24 months, and 5% annual drop-out is considered for sample size calculation. The O'Brien and Fleming alpha spending function is used to allocate the 5% FWER between the IA and FA. Within each panel in Figure 5, vertical solid lines respectively mark the number of events maximising (3) at the IA (left) and FA (right panels), with dashed lines marking the utility maximising number of events at the other pre-planned analysis. The top left panel in Figure 5 shows that, if the interim analysis is planned when 50% of all events expected to be observed over follow-up (information fraction), the design utility (3) does not achieve a maximum when evaluated up to approximately 291 events, which are associated with a trial sample size of approximately 956 patients and 99% cumulative power. The bottom left panel in Figure 5 shows that approximately 350 events maximise IA design utility if the information fraction of the IA is increased to 80%. The panels on the right-hand side of Figure 5 show that a FA utility maximising number of events of approximately 293–310, associated with a sample size of approximately 482–548 study participants, is identified across both IA information fractions. In both cases, the FA critical value is $HR_{CV}(N) \approx 0.79$, with 85% cumulative power. Requiring a larger number of events at FA would achieve higher power but it would also be associated to more modest critical values, greater than 0.8 in this example, with progressively more modest clinical relevance. Overall, the left-hand side panels in Figure 5 show that calibration of sample size with reference to the number of events maximising design utility discourages planning of interim analyses when the information fraction is low. The right-hand side panels in Figure 5 show that utility-based sample size calibration of group sequential design discourages planning overpowered trials, entailing large investment and likely long read-out times when the minimum treatment effect associated to a statistically positive outcome is unlikely to translate into a strong patient benefit in clinical practice.

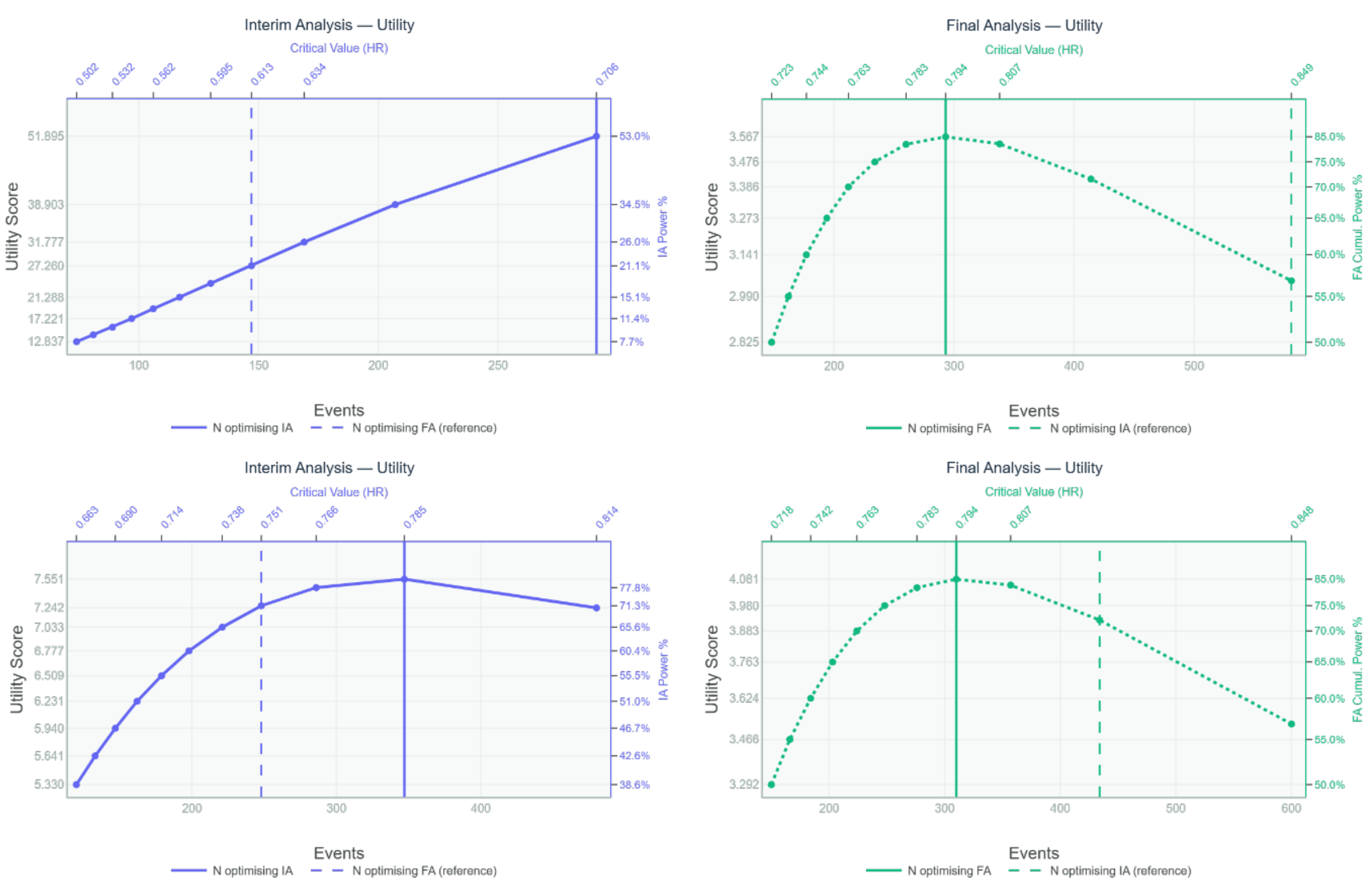

*Figure 5: design utility interim for the interim analysis (left) and final analysis (right) of a group sequential design with time to event primary endpoint, plotted again the number of events expected to be observed in aggregate across the two parallel trial arms. The left plot show that no utility-maximising sample size is reached when the information fraction of the pre-planned interim analysis is 50% (top left). The bottom left plot shows that a utility maximising sample size is reached when the interim analysis information fraction is raised to 80%. Hence, design utility maximisation as a principle for sample size calibration discourages relatively early interim analyses in group sequential designs. The utility maximising sample sizes were identified for the final analysis, regardless of whether an early or later interim was planned. Utility maximisation here discourages designs associated with minimum detectable risk reductions smaller than 20%, which may entail large investment and long read-out times, without necessarily demonstrating practice-changing treatment effects.*

## Discussion

This paper introduced a clinical trial design utility index, applying single attribute utility theory to the optimisation of the trial sample size. This design utility is calculated from standard design parameters, supporting the calibration of trial sample size by balancing statistical power against the minimum treatment effect associated with a statistically positive trial outcome. This approach does not require special software, it does not call for stronger design assumptions compared to current practice, and it is applicable to early as well as to late phase clinical trials. Ownership of the design utility function ultimately rests with the study team, as a tool facilitating but one aspect of the design process. Systematic application of this tool as part of a clear methodology for sample size determination offers an important opportunity for simplification, efficiency, transparency and consistency across trials. Notwithstanding these conceptual advantages, this design utility cannot be taken as normative per se, as additional factors will likely be relevant for determining trial sample size in practice, including patient safety specific to different clinical settings and regulatory considerations.

The illustrative examples provided here empirically demonstrated that unique global maxima were identified for common designs currently in use in all phases of clinical development. This empirical evaluation motivates and justifies the application of the proposed multiplicative utility to the designs presented here. Further analytical characterisations and generalisations, including different utility weights and multilinear utilities, which may accommodate more flexible utility formulations as needed for specific applications beyond those explored here, lie beyond the scope of this paper.

Sample size calibration of a group sequential superiority design with time to event primary efficacy endpoint highlighted two characteristics of the proposed design utility approach. First, interim analyses based on early efficacy results are associated with suboptimal utility. Second, the sample sizes maximising utility respectively at IA and at FA are shown to be different in general. Application of design utility to calibrating sequences of analyses of the same endpoint, or to simultaneous analyses of multiple endpoints, are possible by generalising the univariate design utility proposed here, using multi-attribute utility theory. Pending these further developments, which are beyond the scope of this paper, the univariate design utility proposed here is amenable to applications where the overall trial sample size is optimised with reference to a single pre-specified analysis. From this perspective, the results shown above are encouraging, in that maximisation of design utility for the final analysis provides a clear rationale for sample size calibration, it is robust with respect to the interim analysis information fraction and the power associated to the utility maximising sample size is consistent with current practice in oncology.

**Acknowledgements**

The Authors are grateful to Dr Kristine Broglio for her consultancy on implementation of the Bayesian ORR difference example presented in Section 2. The Authors are also grateful to the Referees and Editors who provided insightful comments on this article.

**Conflict of Interest**

None declared.

**Data availability**

NA

**Funding**

No AstraZeneca funding was used for development of this work.